\documentclass[aps,prl,twocolumn,letterpaper,superscriptaddress, showpacs, amsmath, amssymb]{revtex4}

\usepackage{graphicx}
\usepackage{amsmath,amssymb}
\usepackage{color}

\begin{document}

%%%%%%%%%%%%%%%%%%%%%%%%%%%%%%%%%%%%%%%%%%%%%%%%%%
%%%%%%%%%%%%%%%%%%%%%%%%%%%%%%%%%%%%%%%%%%%%%%%%%%

\title{
TeV--PeV Neutrinos from Low-Power Gamma-Ray Burst Jets inside Stars
}

\author{Kohta Murase}
\affiliation{Hubble Fellow -- Institute for Advanced Study, Princeton, New Jersey 08540, USA}
\author{Kunihito Ioka}
\affiliation{Theory Center, Institute of Particle and Nuclear Studies, KEK, Tsukuba 305-0801, Japan}
\affiliation{Department of Particles and Nuclear Physics, the Graduate University for Advanced Studies (Sokendai), Tsukuba 305-0801, Japan}

\date{17 September 2013}

\begin{abstract}
We study high-energy neutrino production in collimated jets inside progenitors of gamma-ray bursts (GRBs) and supernovae, considering both collimation and internal shocks.  We obtain simple, useful constraints, using the often overlooked point that shock acceleration of particles is ineffective at radiation-mediated shocks.  Classical GRBs may be too powerful to produce high-energy neutrinos inside stars, which is consistent with IceCube nondetections.  We find that ultralong GRBs avoid such constraints and detecting the TeV signal will support giant progenitors.  Predictions for low-power GRB classes including low-luminosity GRBs can be consistent with the astrophysical neutrino background that IceCube may detect, with a spectral steepening around PeV.  The models can be tested with future GRB monitors.   
\end{abstract}

\pacs{95.85.Ry, 97.60.Bw, 98.70.Rz\vspace{-0.3cm}}
% 95.85.Ry Neutrino, muon, pion, and other elementary particles; cosmic rays

\maketitle

%%%%%%%%%%%%%%%%%%%%%%%%%%%%%%%%%%%%%%%%%%%%%%%%%%
%%%%%%%%%%%%%%%%%%%%%%%%%%%%%%%%%%%%%%%%%%%%%%%%%%

%{\it Introduction.---}
%
Long gamma-ray bursts (GRBs) are believed to originate from relativistic jets launched at the death of massive stars.  Associations with core-collapse supernovae (CCSNe) have provided strong evidence for the GRB-CCSN relationship~\cite{grbsn}.  But, there remain many important questions.  What makes the GRB-CCSN connection? How universal is it? What is the central engine and progenitor of GRBs?  How are jets launched and accelerated?  Observationally, it is not easy to probe physics inside a star with photons until the jet breaks out and the photons leave the system.  This is always the case if the jet is ``chocked'' rather than ``successful"~\cite{mw01}; that is, the jet stalls inside the star, where the electromagnetic signal is unobservable.  Such failed GRBs may be much more common than GRBs (whose true rate is $\sim{10}^{-3}$ of that of all CCSNe), and CCSNe driven by mildly relativistic jets may make up a few present of all CCSNe~\cite{llgrb,possiblejet,hn}.

Recent observations suggest interesting diversity in the GRB population.  ``Low-power GRBs'' such as low-luminosity (LL) GRBs~\cite{llgrb,mur+06,llgrbnu} and ultralong (UL) GRBs~\cite{lev+13,gen+13} have longer durations ($\sim{10}^{3}\mbox{--}{10}^{4}$~s) compared to that of classical long GRBs, suggesting different GRB classes and larger progenitors.  While they were largely missed in previous observations, they are important for the total energy budget and the GRB-CCSN connection.  

Neutrinos and gravitational waves (GWs) can present special opportunities to address the above issues.  In particular, IceCube is powerful enough to see high-energy (HE) neutrinos at $\gtrsim1$~TeV~\cite{icecube} and has reported the first detections of cosmic PeV neutrinos~\cite{PeVevents}.  Efficient HE neutrino production inside a star has been proposed assuming shock acceleration of cosmic rays (CRs)~\cite{precursor,slowjet,ha08}, and investigated by a lot of authors, since their detection allows us to study the GRB-CCSN connection~\cite{slowjet,multi}, joint searches with GWs~\cite{gw}, neutrino mixing including the matter effect~\cite{mixing}, the nature of GRB progenitors~\cite{bar+13} and so on.  However, IceCube has not detected neutrinos from GRBs, putting limits on this scenario as well as the classical prompt emission scenario~\cite{grblim,gao+13}.  It also constrains orphan neutrinos from a CCSN~\cite{snlim}. 

In this work, we consider HE neutrino production in a collimated jet inside a star.  Recent developments have revealed that the jet is collimated inside stars rather than ballistic, and becomes slow and cylindrical~\cite{collimation}, where the collimation shock as well as internal shocks occur.  We show that nondetections by IceCube are consistent with theoretical expectations, and more favorable conditions for neutrinos are satisfied in lower-power GRBs such as UL GRBs.  The CR acceleration simply assumed in many studies~\cite{multi,gw,mixing,bar+13} may not be effective in CCSNe and classical GRBs since radiation smoothens the shock structure.
%Such less luminous transients, which have been recently discovered, hold a key to the engine and progenitor of GRBs, and our study will be useful for many studies that simply assumed CR acceleration~\cite{multi,gw,mixing,bar+13}.  
Also, the low-power GRBs could significantly contribute to the extragalactic neutrino background (ENB), which consists of the sum of contributions from sources at various redshifts and may have been seen by IceCube~\cite{PeVevents}.  Throughout this work, we use $Q_x=Q/10^x$ in CGS unit with cosmological parameters of $H_0=71~{\rm km}~{\rm s}^{-1}~{\rm Mpc}^{-1}$, $\Omega_m=0.3$, and $\Omega_\Lambda=0.7$.

%%%%%%%%%%%%%%%%%%%%%%%%%%%%%%%%%%%%%%%%%%%%%%%%%%
%%%%%%%%%%%%%%%%%%%%%%%%%%%%%%%%%%%%%%%%%%%%%%%%%%

{\it Jet propagation in a star.---}
To make a GRB, a jet must penetrate the progenitor successfully.  The jet dynamics is governed by the jet head, cocoon and collimation~\cite{collimation}.  The jet is decelerated by a reverse shock while a forward shock is formed in the stellar envelope.  The jet head, which is the shocked region between the two shocks, is controlled by the ram pressure balance between the reverse-shocked jet and forward-shocked envelope~\cite{mw01,mat03}.  This shocked region is so hot to expand sideways to form a cocoon.  For a given initial opening angle $\theta_j$, if the absolute jet luminosity $L_{j0}$ is low enough and/or the ambient density $\varrho_a$ is high enough, the hydrodynamic jet is collimated by the cocoon pressure via collimation shocks (see Fig.~1).  At time $t$, the (collimation-)shocked jet becomes cylindrical through the collimation shock at~\cite{collimation} 
\begin{equation}
r_{\rm cs}\approx4.1\times{10}^{9}~{\rm cm}~t^{2/5}L_{j0,52}^{3/10}{(\theta_{j}/0.2)}^{-1/5}\varrho_{a,4}^{-3/10},
\end{equation}
beyond which the cylindrical, collimated flow has a constant Lorentz factor (with $\Gamma_{\rm cj}\approx\theta_j^{-1}$) because of the flux conservation.  The subsequent jet head position $r_h$ is 
\begin{equation}
r_{h}\approx8.0\times{10}^{9}~{\rm cm}~t^{3/5}L_{j0,52}^{1/5}{(\theta_{j}/0.2)}^{-4/5}\varrho_{a,4}^{-1/5}.
\end{equation}
Even if the jet achieves $\Gamma\gg\Gamma_{\rm cj}$ in the star, $\Gamma_{\rm cj}\approx5{(\theta_j/0.2)}^{-1}$ implies that the collimated jet is radiation dominated.  The jet breakout time $t_{\rm bo}$ is determined by $r_h(t_{\rm bo})=R_*$, where $R_*$ is the progenitor radius.

%%%%%%%%%%%%%%%%%%%%%%%%%%%%%%%%%%
\begin{figure}[t]
\includegraphics[width=3.00in]{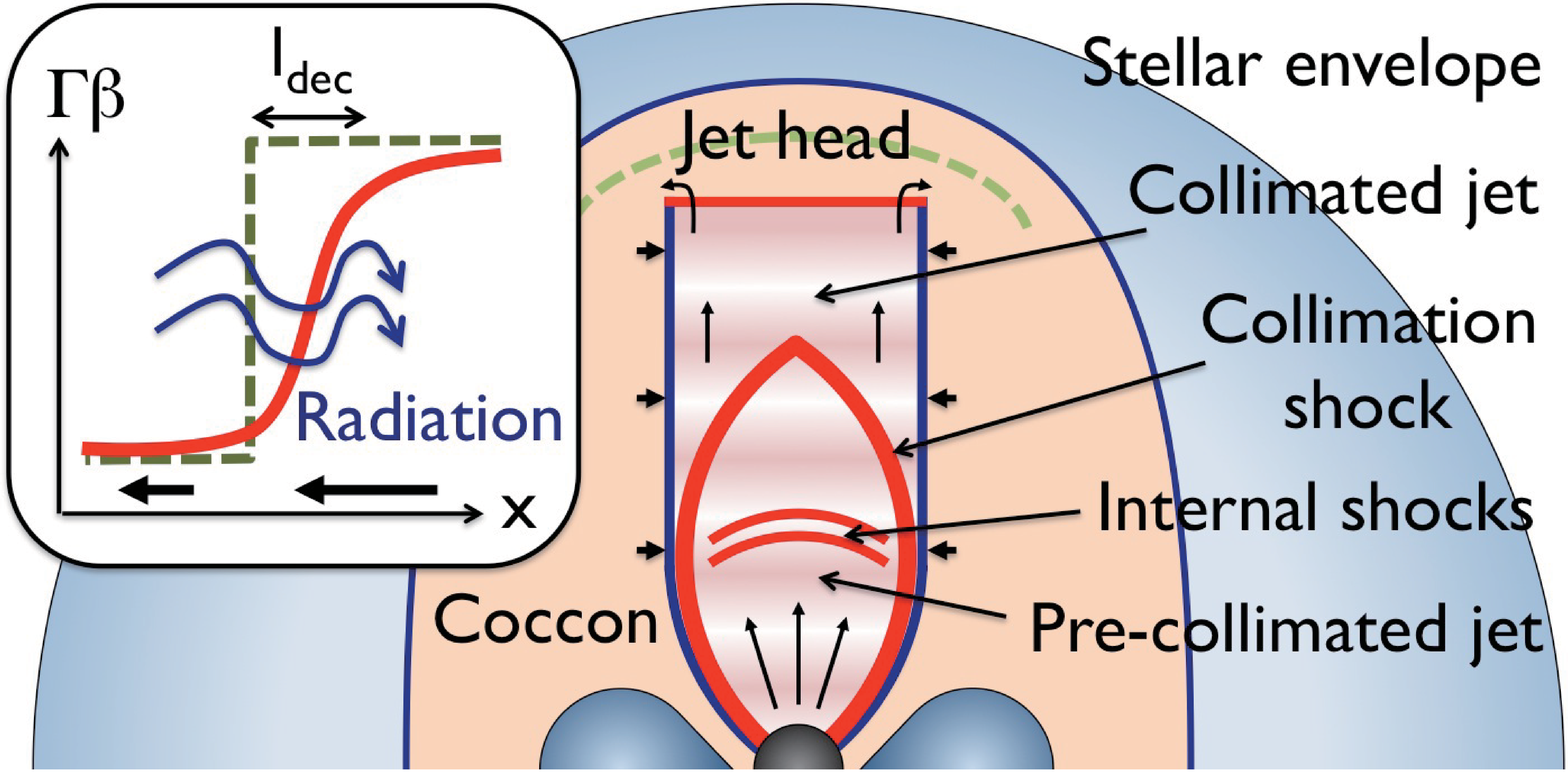}
\caption{
The schematic picture of a collimated GRB jet inside a progenitor.  CR acceleration and HE neutrino production may happen at collimation and internal shocks.  The picture of the radiation-mediated shock is also shown.
}
\vspace{-1.\baselineskip}
\end{figure}
%%%%%%%%%%%%%%%%%%%%%%%%%%%%%%%%%%

The progenitor of long GRBs has been widely believed to be a star without an envelope, such as Wolf-Rayet (WR) stars with $R_*\sim0.6\mbox{--}3R_{\odot}$~\cite{cro07}.  Let us approximate the density profile to be $\varrho_a=(3-\alpha)M_*{(r/R_*)}^{-\alpha}/(4\pi R_*^3)$ ($\alpha\sim1.5\mbox{--}3$), where $M_*$ is the progenitor mass~\cite{mm99}.  Then, taking $\alpha=2.5$, we obtain $r_{\rm cs}\approx1.6\times{10}^{9}~{\rm cm}~t_1^{8/5}L_{0,52}^{6/5}{(\theta_j/0.2)}^{8/5}{(M_{*}/20~M_{\odot})}^{-6/5}R_{*,11}^{3/5}$ and $r_h\approx5.4\times{10}^{10}~{\rm cm}~t_1^{6/5}L_{0,52}^{2/5}{(\theta_j/0.2)}^{-4/5}{(M_{*}/20~M_{\odot})}^{-2/5}\\R_{*,11}^{1/5}$~\cite{collimation}, where $L_0=4L_{0j}/\theta_j^2$ is the isotropic total jet luminosity.  The GRB jet is successful if $t_{\rm bo}\approx17~{\rm s}~L_{0,52}^{-1/3}{(\theta_j/0.2)}^{2/3}{(M_{*}/20~M_{\odot})}^{1/3}R_{*,11}^{2/3}$ is shorter than the jet duration $t_{\rm dur}$.  With $t_{\rm dur}\sim30$~s, we typically expect $r_{\rm cs}\sim{10}^{10}~{\rm cm}$ for classical GRBs~\cite{comment1}.

The comoving proton density in the collimated jet is $n_{\rm cj}\approx L_0/(4\pi r_{\rm cs}^2 \Gamma_{\rm cj}\eta m_pc^3)=L/(4\pi r_{\rm cs}^2\Gamma_{\rm cj}\Gamma m_pc^3)\simeq3.5\times{10}^{20}~{\rm cm}^{-3}~L_{52}r_{\rm cs,10}^{-2}\Gamma_{2}^{-1}(5/\Gamma_{\rm cj})$. 
Here, $L=(\Gamma/\eta)L_{0}$, $L$ is the isotropic kinetic luminosity, and $\eta$ is the maximum Lorentz factor.  The density in the precollimated jet at the collimation or internal shock radius $r_s$ is $n_j\approx L/(4 \pi r_{s}^2\Gamma^2m_pc^3)\simeq1.8\times{10}^{19}~{\rm cm}^{-3}~L_{52}r_{s,10}^{-2}\Gamma_{2}^{-2}$, which is lower than $n_{\rm cj}$ due to $\Gamma\gg\Gamma_{\rm cj}$.  This quantity is relevant in discussions below.  Note that inhomogeneities in the jet lead to internal shocks, where the Lorentz factor can be higher ($\Gamma_r$) and lower ($\Gamma_s$) than $\Gamma\approx\sqrt{\Gamma_r\Gamma_s}$. 

%%%%%%%%%%%%%%%%%%%%%%%%%%%%%%%%%%%%%%%%%%%%%%%%%%
%%%%%%%%%%%%%%%%%%%%%%%%%%%%%%%%%%%%%%%%%%%%%%%%%%

{\it Radiation constraints.---}
Efficient CR acceleration at internal shocks and the jet head has been suggested, since plasma time scales are typically shorter than any elastic or inelastic collision time scale~\cite{precursor,slowjet,ha08}.  However, in the context of HE neutrinos from GRBs, it has often been overlooked that shocks deep inside a star may be radiation mediated~\cite{lb08}.  At such shocks, photons produced in the downstream diffuse into the upstream and interact with electrons (plus pairs).  Then the upstream proton flow should be decelerated by photons via coupling between thermal electrons and protons~\cite{rms1}.  As a result (see Fig.~1), one no longer expects a strong shock jump (although a weak subshock may exist~\cite{rms2}), unlike the usual collisionless shock, and the shock width is determined by the deceleration scale $l_{\rm dec}\approx{(n_u\sigma_Ty_\pm)}^{-1}\simeq1.5\times{10}^5~{\rm cm}~n_{u,19}^{-1}y_\pm^{-1}$ when the comoving size of the upstream flow $l_u$ is longer than $l_{\rm dec}$.  Here $n_u$ is the upstream proton density, and $y_\pm(\geq1)$ is the possible effect of pairs entrained or produced by the shock~\cite{csn}.  

In the conventional shock acceleration, CRs are injected at quasithermal energies~\cite{dsa}.  The Larmor radius of CRs with $\sim\Gamma_{\rm rel}^2m_pc^2$ is $r_L^u\sim\Gamma_{\rm rel}^2m_pc^2/(eB)\simeq3.8\times{10}^{-3}~{\rm cm}~\epsilon_B^{-1/2}L_{0,52}^{-1/2}r_{s,10}\Gamma_{2}\Gamma_{\rm rel}^2$, where $B$ is the magnetic field, $\Gamma_{\rm rel}$ is the relative Lorentz factor and $\epsilon_B\equiv L_B/L_0$~\cite{comment2}.  
If the velocity jump of the flow is small over $r_L^{u}$, the CR acceleration is inefficient.  For $l_{\rm dec}\ll l_u$, since significant deceleration occurs over $\sim l_{\rm dec}$, including the immediate upstream~\cite{rms1,rms2}, CRs with $r_L^u\ll l_{\rm dec}$ do not feel the strong compression and the shock acceleration will be suppressed~\cite{lb08,comment3,commentnpc}.  CRs are expected when photons readily escape from the system and the shock becomes radiation unmediated, which occurs when $l_u\lesssim l_{\rm dec}$~\cite{csn,rsg}.  
Regarding this as a reasonably necessary condition for the CR acceleration, we have
\begin{equation}
\tau_T^{u}=n_u\sigma_Tl_u\lesssim {\rm min}[1,0.1C^{-1}\Gamma_{\rm rel}],
\end{equation}
where $C=1+2\ln \Gamma_{\rm rel}^2$ is the possible effect by pair production~\cite{rms2}, although it may be small when photons start to escape.  Since the detailed pair-production effect is uncertain, $\tau_T^u\lesssim1$ gives us a conservative bound.

Applying Eq.~(3) to the collimation shock~\cite{comment4}, the radiation constraint for the CR acceleration is
\begin{equation}
L_{52}r_{\rm cs,10}\Gamma_2^{-3}\lesssim5.7\times{10}^{-4}~{\rm min}[1,0.01C_{1}^{-1}\Gamma_{\rm rel}],
\end{equation}
where $n_u=n_j$, $l_u\approx r_{\rm cs}/\Gamma$, and $\Gamma_{\rm rel}\approx(\Gamma/\Gamma_{\rm cj}+\Gamma_{\rm cj}/\Gamma)/2$ are used.  As shown in Fig.~2, it is difficult to expect CRs and HE neutrinos from the collimation shock for classical GRBs.  We note that the termination shock at the jet head and internal shocks in the collimated jet are less favorable for the CR acceleration than the collimation shock since $n_{\rm cj}\gg n_{j}$ and $\Gamma_{\rm cj}\ll\Gamma$.

%%%%%%%%%%%%%%%%%%%%%%%%%%%%%%%%%%
\begin{figure}[t]
\includegraphics[width=3.00in]{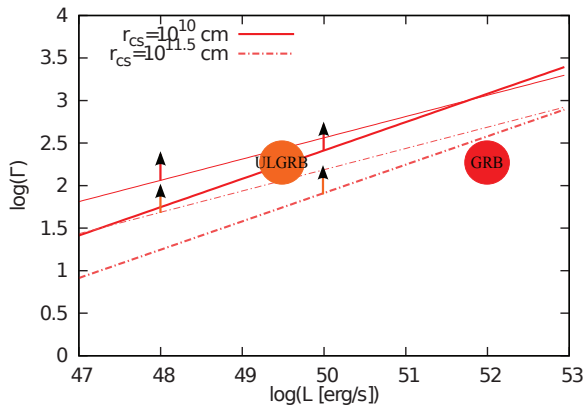}
\caption{
Lower limits on $\Gamma$ for given $L$, above which CRs and HE neutrinos can be expected from the collimation shock (without mediation by radiation) at $r_{\rm cs}$.  Thick (thin) curves represent cases without (with) the possible pair effect with the approximation of $C\simeq10$.  Typical parameters of classical GRBs and UL GRBs are depicted.   
\label{fig:cs}
}
\vspace{-1.\baselineskip}
\end{figure}
%%%%%%%%%%%%%%%%%%%%%%%%%%%%%%%%%%

We can also apply Eq.~(3) to internal shocks in the precollimated jet, which have been considered in the literature~\cite{precursor,slowjet}.  Internal shocks may occur above $r_{\rm is}\approx2\Gamma_s^2c\delta t\simeq3.0\times{10}^{10}~{\rm cm}~\Gamma_{s,1.5}^2{\delta t}_{-3}$, and the relative Lorentz factor between the rapid and merged shells is $\Gamma_{\rm rel}\approx(\Gamma_r/\Gamma+\Gamma/\Gamma_r)/2$, which may lead to the upstream density in the rapid shell $\sim n_j/\Gamma_{\rm rel}$.  Using $l_u\approx r_{\rm is}/\Gamma_r\sim l/\Gamma_{\rm rel}$, we get $\tau_T=n_{j}\sigma_Tl\lesssim{\rm min}[\Gamma_{\rm rel}^2,0.1C^{-1}\Gamma_{\rm rel}^3]$ or
\begin{equation}
L_{52}r_{\rm is,10}\Gamma_2^{-3}\lesssim5.7\times{10}^{-3}~{\rm min}[\Gamma_{\rm rel,0.5}^2,0.32C_{1}^{-1}\Gamma_{\rm rel,0.5}^3].
\end{equation} 
As shown in Fig.~3, unless $\Gamma\gtrsim{10}^3$, it seems difficult to expect CRs and HE neutrinos for high-power jets inside WR-like progenitors (where $r_{\rm is}\lesssim r_{\rm cs}\sim{10}^{10}$~cm).  Note that although the constraint is relevant for shocks deep inside the stars we here consider, CRs may be expected around the photosphere $\tau_T\sim1\mbox{--}10$~\cite{mur08}, as assumed in the dissipative photosphere scenario~\cite{photosphere}.

The radiation constraint is useful for the slow-jet model in which CCSNe are driven by mildly relativistic jets with $\Gamma\sim2\mbox{--}10$~\cite{slowjet}.  Interestingly, it is complementary to observations.  IceCube placed upper limits on $\Gamma$ for given ${\mathcal E}_j=2L_jt_{\rm dur}$ model dependently~\cite{snlim}, and its upper limit is shown in Fig.~3 after converting ${\mathcal E}_j$ to $L$.  We may expect HE neutrinos from sufficiently low-power 
jets with $L\lesssim{10}^{45.5}\mbox{--}{10}^{47}~{\rm erg}~{\rm s}^{-1}$ for WR-like progenitors.

%%%%%%%%%%%%%%%%%%%%%%%%%%%%%%%%%%
\begin{figure}[t]
\includegraphics[width=3.15in]{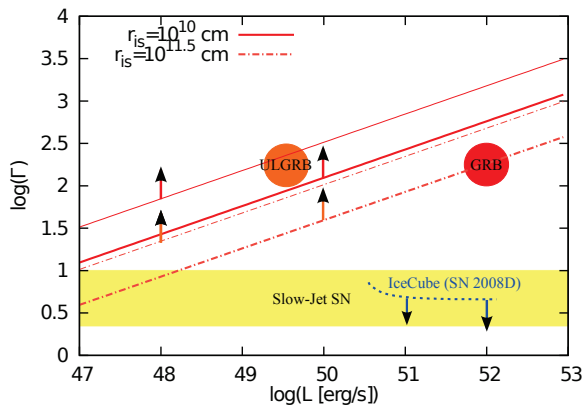}
\caption{
The same as Fig.~2, but for the internal shock (with $\Gamma_{\rm rel}=3$) at $r_{\rm is}$.  The IceCube upper limit on the slow-jet model for SN 2008D~\cite{snlim} is also shown for $t_{\rm dur}=100$~s.
}
\vspace{-1.\baselineskip}
\end{figure}
%%%%%%%%%%%%%%%%%%%%%%%%%%%%%%%%%%

%%%%%%%%%%%%%%%%%%%%%%%%%%%%%%%%%%%%%%%%%%%%%%%%%%
%%%%%%%%%%%%%%%%%%%%%%%%%%%%%%%%%%%%%%%%%%%%%%%%%%

{\it Ultralong GRBs.---}
As seen above, efficient CR acceleration may not occur in high-power jets inside WR-like progenitors.  However, the situation is different for less-power GRBs such as UL GRBs~\cite{lev+13,gen+13} and LL GRBs~\cite{llgrb}.  In particular, UL GRBs are as energetic and possibly common as classical GRBs~\cite{lev+13}.  Their lower-luminosity $L_\gamma\sim{10}^{49}\mbox{--}{10}^{50}~{\rm erg}~{\rm s}^{-1}$ and longer duration $t_{\rm dur}\sim{10}^{4}$~s suggest bigger progenitors like blue supergiants (BSGs) with $R_*\sim{10}^{12}\mbox{--}{10}^{13}$~cm~\cite{gen+13,bsg}. 

Assuming a stellar envelope with $\varrho_a(r)={10}^2~{\rm g}~{\rm cm}^{-3}~\varrho_{\rm BSG}~r_{10}^{-2}$~\cite{envelope}, with Eqs.~(1) and (2), we obtain $r_{\rm cs}\simeq1.4\times{10}^{11}~{\rm cm}~t_4L_{0,49.5}^{3/4}(\theta_j/0.2)\varrho_{\rm BSG}^{-3/4}$, $r_h\simeq1.0\times{10}^{13}~{\rm cm}~t_4L_{0,49.5}^{1/3}{(\theta_j/0.2)}^{-2/3}\varrho_{\rm BSG}^{-1/3}$, and $t_{\rm bo}\simeq9800~{\rm s}~L_{0,49.5}^{-1/3}{(\theta_j/0.2)}^{2/3}\varrho_{\rm BSG}^{1/3}R_{*,13}$ (comparable to $t_{\rm dur}$).  We may typically expect the collimation shock at $r_{\rm cs}\sim{10}^{11.5}$~cm.

Interestingly, thanks to lower powers and larger shock radii, the Thomson optical depth is low even in a star ($\tau_T\approx0.12~L_{49.5}r_{s,11.5}^{-1}\Gamma_{2}^{-3}$), allowing the CR acceleration and HE neutrino production, as indicated in Figs.~2 and 3.  The jet may be sufficiently accelerated by $r_{\rm cs}$~\cite{comment5}, while should slow down to $\Gamma_{\rm cj}$ after the collimation. 

Once CRs are accelerated inside a star, the CR power is lost to meson production via the $p\gamma$ reaction with target photons or the $pp$ reaction with target nucleons, leading to {\it precursor or orphan} neutrinos.  We consider two possibilities: HE neutrinos from CRs accelerated at the collimation shock (CS) and HE neutrinos from CRs accelerated at the internal shock in the precollimated jet (IS). 

In the CS scenario, CRs are conveyed in the collimated jet with $\Gamma_{\rm cj}$ and completely depleted during the advection for $R_{\rm adv}/c\approx{\rm min}[R_{*},r_h(t_{\rm dur})]/c$.  Using the photon temperature $kT_{\rm cj}\approx0.70~{\rm keV}~L_{0,49.5}^{1/4}r_{\rm cs,11.5}^{-1/2}{(\Gamma_{\rm cj}/5)}^{-1/2}$ in the collimated jet and $\hat{\sigma}_{p\gamma}\approx0.58\times{10}^{-28}~{\rm cm}^2$, we obtain the high $p\gamma$ efficiency $f_{p\gamma}^{\rm cj}\approx n_\gamma^{\rm cj}\hat{\sigma}_{p\gamma}(R_{\rm adv}/\Gamma_{\rm cj})\simeq1.2\times{10}^6~L_{0,49.5}^{3/4}r_{\rm cs,11.5}^{-3/2}{(\Gamma_{\rm cj}/5)}^{-5/2}R_{\rm adv,13}\gg1$.  The $pp$ efficiency is also high since $f_{pp}^{\rm cj}\approx n_{\rm cj}\hat{\sigma}_{pp}(R_{\rm adv}/\Gamma_{\rm cj})\simeq56~L_{49.5}r_{\rm cs,11.5}^{-2}\Gamma_{2}^{-1}{(\Gamma_{\rm cj}/5)}^{-2}R_{\rm adv,13}\gg1$.  CRs are depleted essentially in the entire energy range, so the system is ``calorimetric'' and HE neutrinos are unavoidable.  Since the formation of collimation shocks is also quite common for relativistic jets inside stars, HE neutrinos from UL GRBs can be used as signatures of jets in big progenitors.  
Note that due to copious target photons, the maximum energy in the acceleration zone $\varepsilon_p^M$ is limited by the $p\gamma$ reaction.  By comparing the acceleration time $t_{\rm acc}\approx\varepsilon_p/(eBc)$ to the $p\gamma$ cooling time $t_{p\gamma}\approx1/(n_{\gamma}^{\rm cj}\hat{\sigma}_{p\gamma}c)$, we obtain $\varepsilon_p^M\simeq1.5\times{10}^{6}~{\rm GeV}~B_{6.5}L_{0,49.5}^{-3/4}r_{\rm cs,11.5}^{3/2}{(\Gamma_{\rm cj}/5)}^{3/2}$.    

In the IS scenario, during the dynamical time, CRs mainly interact with photons escaping back from the collimated jet.  Using the photon density $n_\gamma^{j}\approx(\Gamma/2\Gamma_{\rm cj})(f_{\rm esc}n_\gamma^{\rm cj})$ [where $f_{\rm esc}\sim{(n_{\rm cj}\sigma_Tr_{\rm cs}/\Gamma_{\rm cj})}^{-1}$ is the escape fraction], which is boosted by $\Gamma_{\rm rel}\sim\Gamma/2\Gamma_{\rm cj}$, we have $f_{p\gamma}^j\approx(\Gamma/2\Gamma_{\rm cj})(f_{\rm esc}n_\gamma^{\rm cj})\hat{\sigma}_{p\gamma}(r_{\rm is}/\Gamma)\gg1$, so HE CRs with $f_{p\gamma}^j\gtrsim1$ are depleted as in the CS scenario.  One has $f_{p\gamma}^j\sim1$ near the typical $p\gamma$ threshold energy, $\varepsilon_p^{\rm th}\simeq1.7\times{10}^{3}~{\rm GeV}~L_{0,49.5}^{-1/4}r_{\rm cs,11.5}^{1/2}\Gamma_2^{-1}{(\Gamma_{\rm cj}/5)}^{3/2}$.  On the other hand, since $n_j$ is small, the $pp$ efficiency is too low to be relevant.  As in the CS scenario, $\varepsilon_p^M$ is limited by the $p\gamma$ process, leading to $\varepsilon_p^M\approx eB/(n_\gamma^{j}\hat{\sigma}_{p\gamma})$. 

We calculate neutrino spectra, using the numerical code developed in Refs.~\cite{mur08,mur07,mur+06}, where $p\gamma/pp$ reactions and relevant cooling processes are considered in detail.  Note that we consistently evaluate $\varepsilon_p^M$ by comparing $t_{\rm acc}$ with all relevant competing time scales.  We get $\varepsilon_p^M\sim{10}^{6.3}$~GeV and $\varepsilon_p^M\sim{10}^{6.1}$~GeV in the CS and IS scenarios, respectively.  Then, we calculate depletion of CRs and neutrino spectra, assuming a CR spectrum of $\varepsilon_p^{-2}{e}^{-\varepsilon_p/\varepsilon_p^M}$.  The parameters are shown in Fig.~4.  We assume $\epsilon_B=1$ in the IS scenario, while $L_B^{\rm cj}={10}^{-2}L_0$ in the CS scenario since the collimated jet is radiation dominated and its magnetic luminosity would be smaller than the kinetic luminosity, but key results are not sensitive when the meson synchrotron cooling is subdominant~(cf. Ref.~\cite{slowjet}).

%%%%%%%%%%%%%%%%%%%%%%%%%%%%%%%%%%
\begin{figure}[t]
\includegraphics[width=3.15in]{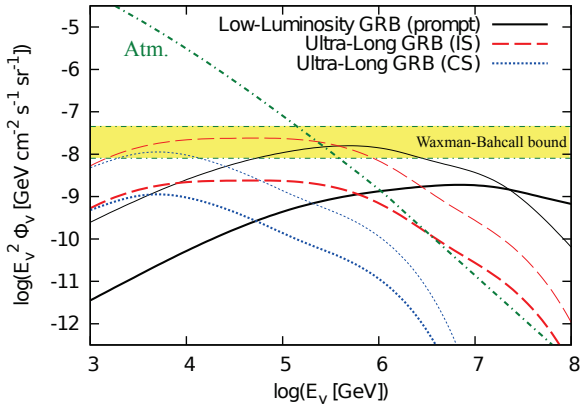}
\caption{The cumulative neutrino backgrounds from UL GRBs and LL GRBs.  For UL GRBs, we use $r_s={10}^{11.5}$~cm, $\Gamma_{\rm cj}=5$, $kT_{\rm cj}\simeq0.70$~keV, $\Gamma=100$ and $L={10}^{49}~{\rm erg}~{\rm s}^{-1}$.  The CR energy generation rate is set to $\xi_{\rm acc}{\mathcal E}_{\gamma}^{\rm iso}\rho={10}^{53}~{\rm erg}~{\rm Gpc}^{-3}~{\rm yr}^{-1}$, with $f_{\rm cho}=1$ (thick) and $f_{\rm cho}=10$ (thin).  For comparison, predictions for prompt emission from LL GRBs (with $\rho=500~{\rm Gpc}^{-3}~{\rm yr}^{-1}$ and $\xi_{\rm acc}=10$) are taken from Ref.~\cite{mur+06} for $\Gamma=10$ (thick) and $\Gamma=5$ (thin).  For redshift evolution, the GRB3 model is assumed~\cite{mur07}.  The atmospheric background~\cite{atm} is also shown.  Note that IceCube suggests $E_\nu^2\Phi_\nu\sim{\rm a~few}\times{10}^{-8}~{\rm GeV}~{\rm cm}^{-2}~{\rm s}^{-1}~{\rm sr}^{-1}$~\cite{PeVevents}, which is compatible with the original Waxman-Bahcall bound~\cite{wb}.
}
\vspace{-1.\baselineskip}
\end{figure}
%%%%%%%%%%%%%%%%%%%%%%%%%%%%%%%%%%

The expected number of neutrino events from a burst at $z=0.1$ is at most $\sim1$, so aggregating many bursts is important.  Alhough it is hard for current satellites to find many low-power GRBs, we can in principle test the scenarios by stacking neutrino signals from $\gtrsim100$ UL GRBs at $z\sim1$, which are detectable by all-sky monitors with sensitivities better than \textit{Swift}.
   
To demonstrate their neutrino spectra and contributions, we numerically calculate the total ENB~\cite{mur07}, which is consistent with the following analytical formula~\cite{mur+06,wb}:
\begin{eqnarray}
E_\nu^2\Phi_\nu&\sim&\frac{c}{4\pi H_0}\frac{3}{8}f_{\rm sup}{\rm min}[1,f_{p\gamma}]E_p^2\frac{dN_p^{\rm iso}}{dE_p}\rho f_zf_{\rm cho}\\
&\sim&4\times{10}^{-8}~{\rm GeV}~{\rm cm}^{-2}~{\rm s}^{-1}~{\rm sr}^{-1}~(f_{\rm cho}\xi_{\rm acc}/10)f_{\rm sup}\nonumber\\
&\times&{\rm min}[1,f_{p\gamma}]({\mathcal E}_{\gamma}^{\rm iso}\rho/{10}^{53}~{\rm erg}~{\rm Gpc}^{-3}~{\rm yr}^{-1})(f_z/3),\nonumber
\end{eqnarray}
where $f_z$ is the evolution factor~\cite{wb}, $f_{\rm sup}$ is the suppression factor due to the meson and muon cooling~\cite{mur08}, $\xi_{\rm acc}$ is the CR loading parameter~\cite{mur+06}, and $f_{\rm cho}$ is the fraction of failed GRBs compared to successful GRBs.  Here, $\rho$ is the local rate that is $\sim1~{\rm Gpc}^{-3}~{\rm yr}^{-1}$ for GRBs and UL GRBs~\cite{lev+13} (but see Ref.~\cite{gen+13}) while $\sim{10}^{2}\mbox{--}{10}^{3}~{\rm Gpc}^{-3}~{\rm yr}^{-1}$ for LL GRBs~\cite{llgrb}.  

Results are shown in Fig.~4, where we see that the ENB flux from successful UL GRB jets inside stars may be $\sim{10}^{-9}~{\rm GeV}~{\rm cm}^{-2}~{\rm s}^{-1}~{\rm sr}^{-1}$.  If failed UL GRBs are $\gtrsim10$ times more common, $\sim{10}^{-8}~{\rm GeV}~{\rm cm}^{-2}~{\rm s}^{-1}~{\rm sr}^{-1}$ may even be achieved.  Although the uncertainty in $\rho$ is large, contributions from LL GRBs~\cite{mur+06,llgrbnu,csn} and/or failed UL GRBs can be compatible to the ENB that IceCube may start to observe~\cite{PeVevents}.  The spectral steepening is also expected. In particular, in the IS scenario, the meson radiative cooling or the cutoff from the proton maximum energy can lead to a break around PeV.  In addition, for choked jets in BSGs, the cutoff at $\gtrsim1$~PeV is possible due to neutrino absorption in the envelope if $r_h\gtrsim5\times{10}^{12}$~cm.  In the CS scenario, strong meson cooling leads to a break at $\lesssim10$~TeV, so we mainly expect multi-TeV neutrinos.   

%%%%%%%%%%%%%%%%%%%%%%%%%%%%%%%%%%%%%%%%%%%%%%%%%%
%%%%%%%%%%%%%%%%%%%%%%%%%%%%%%%%%%%%%%%%%%%%%%%%%%

{\it Summary and discussion.---}
We derived general constraints on HE neutrino production in GRB jets inside stars, based on the point that the shock acceleration is inefficient at radiation-mediated shocks.  They are complementary to observational upper limits, and current nondetections of precursor (orphan) neutrinos from GRBs (CCSNe) are consistent with theoretical expectations.  Our work is encouraging and useful for the literature on the GRB-CCSN connection~\cite{multi}, joint searches with GWs~\cite{gw}, and neutrino mixing~\cite{mixing}.

We showed that more favorable conditions for HE neutrino production are satisfied in low-power GRBs such as UL GRBs especially if they originate from bigger progenitors like BSGs.  The formation of collimation shocks is naturally expected, so TeV neutrinos are useful as a smoking gun of jet physics that cannot be probed with photons, and will also support the idea of BSG-like progenitors.  We stress the importance of stacking such less luminous transients with next-generation all-sky monitors like SVOM, Lobster, WF-MAXI and HiZ-Gundam.

Internal shocks in a precollimated jet could extend the ENB to PeV energies, which may give an important contribution if failed UL GRBs are $\gtrsim10$ times more common.  Note that the neutrino production site considered in this work is different from the prompt emission site.  Since low-power GRBs may be largely missed, even if their successful jets give $\sim{10}^{-9}~{\rm GeV}~{\rm cm}^{-2}~{\rm s}^{-1}~{\rm sr}^{-1}$, the results may not contradict with nondetections of ``prompt'' neutrinos from classical GRBs, which placed $\lesssim{10}^{-9}~{\rm GeV}~{\rm cm}^{-2}~{\rm s}^{-1}~{\rm sr}^{-1}$~\cite{grblim}.  LL GRBs can give $\sim{10}^{-8}~{\rm GeV}~{\rm cm}^{-2}~{\rm s}^{-1}~{\rm sr}^{-1}$, as predicted in Refs.~\cite{csn,mur+06,llgrbnu}.  They are distinct from classical GRBs and they may be more baryon rich~\cite{alternate}.  Since the uncertainty in $\rho$ is large, revealing these transients, which have been largely missed so far, is important to test the models.

%%%%%%%%%%%%%%%%%%%%%%%%%%%%%%%%%%%%%%%%%%%%%%%%%%
%%%%%%%%%%%%%%%%%%%%%%%%%%%%%%%%%%%%%%%%%%%%%%%%%%

\medskip
{\it Acknowledgments.---}
K. M. thanks Omer Bromberg, Boaz Katz, Peter M\'esz\'aros, Tsvi Piran and Eli Waxman for useful discussions and acknowledges the CCAPP workshop, Revealing Deaths of Massive Stars with GeV-TeV Neutrinos.  This work is supported by NASA through Hubble Fellowship, Grant No. 51310.01 awarded by the STScI, which is operated by the Association of Universities for Research in Astronomy, Inc., for NASA, under Contract No. NAS 5-26555 (K. M.) and the Grants-in-Aid for Scientific Research No. 24103006, No. 24000004, No. 22244030 of MEXT and JSPS (K. I.).

%%%%%%%%%%%%%%%%%%%%%%%%%%%%%%%%%%%%%%%%%%%%%%%%%%
%%%%%%%%%%%%%%%%%%%%%%%%%%%%%%%%%%%%%%%%%%%%%%%%%%


\begin{thebibliography}{99}
\bibitem{comment1}
Simulations are needed to study details after $t_{\rm bo}$.
\bibitem{comment2}
In principle, if the turbulent magnetic field is weak and/or the coherence length is small, the CR mean free path can be longer than $l_{\rm dec}$, where CR acceleration is possible but slow and other losses would become relevant.
\bibitem{comment3}
CRs with $\varepsilon_p\gtrsim eBl_{\rm dec}$ may feel the strong compression since their Larmor radius is larger than $l_{\rm dec}$.  But we do not expect efficient CR acceleration since we typically find $\varepsilon_p^M\lesssim eBl_{\rm dec}$ (mainly due to $n_\gamma\hat{\sigma}_{p\gamma}>n_uy_\pm\sigma_{T}$).  Here $\varepsilon_p$ is the proton energy in the comoving frame.
\bibitem{commentnpc}
If neutrons are loaded, the neutron-proton-converter acceleration occurs~\cite{npc}, although it is slow.
\bibitem{comment4}
The shock geometry may play some roles, which is beyond the scope of this work.
\bibitem{comment5}
For example, if it is magnetically accelerated as $\Gamma\approx\eta(r/r_0)^{1/3}$~(e.g., Ref.~\cite{magnetic}), $\Gamma\sim100$ is achieved at $r\sim{10}^{11.5}$~cm for $\eta={10}^{2.5}$ and $r_0={10}^{8}$~cm.
\end{thebibliography}
\end{document}